\documentclass[structabstract]{aa} 
\usepackage{tabularx}                                  
\usepackage{enumerate}
\usepackage{graphicx}
\usepackage{txfonts}
\usepackage{multirow}
\usepackage[sort]{natbib}

\begin{document}

   \title{The galaxy population of the complex cluster system Abell 3921\thanks{Tab. 5 is only available in electronic form at the CDS via anonymous ftp to cdsarc.u-strasbg.fr (130.79.128.5) or via http://cdsweb.u-strasbg.fr/cgi-bin/qcat?J/A+A/}}

   \author{Florian Pranger\inst{1}\fnmsep\thanks{\email{florian.pranger@uibk.ac.at}}
          \and Asmus B\"{o}hm\inst{1}
          \and Chiara Ferrari\inst{2}
          \and Antonaldo Diaferio\inst{3}
          \and Richard Hunstead\inst{4}
          \and Sophie Maurogordato\inst{2}
          \and Christophe Benoist\inst{2}
          \and Jarle Brinchmann\inst{5}
          \and Sabine Schindler\inst{1}
          }

   \institute{
Institute for Astro- and Particle Physics, University of Innsbruck, Technikerstr. 25/8, A-6020 Innsbruck, Austria
\and Laboratoire Lagrange, UMR7293, Universit\'{e} de Nice Sophia Antipolis, CNRS, Observatoire de la C\^{o}te d'Azur, 06300, Nice, France
\and Dipartimento di Fisica, Universit\`{a} degli Studi di Torino, Via P. Giuria 1, I-10125 Torino, Italy; Istituto Nazionale di Fisica Nucleare (INFN), Sezione di Torino, Via P. Giuria 1, I-10125 Torino, Italy
\and School of Physics, University of Sydney, NSW 2006, Australia
\and  Leiden Observatory, Leiden University, 2300 RA Leiden, The Netherlands
}          
   \date{Received \today; accepted ???}
 
  \abstract
  {We present a spectrophotometric analysis of the galaxy population in the area of the merging cluster Abell 3921 at z=0.093.}   
   {We investigate the impact of the complex cluster environment on galaxy properties such as morphology or star formation rate.}
   {We combine multi-object spectroscopy from the two-degree field (2dF) spectrograph with optical imaging taken with the ESO Wide Field Imager. We carried out a redshift analysis and determine cluster velocity dispersions using biweight statistics. Applying a Dressler-Shectman test we sought evidence of cluster substructure. Cluster and field galaxies were investigated with respect to [OII] and H$\alpha$ equivalent width, star formation rate, and morphological descriptors, such as concentration index and Gini coefficient. We studied these cluster galaxy properties as a function of clustercentric distance and investigated the spatial distribution of various galaxy types.}
   {Applying the Dressler-Shectman test, we find a third component (A3921-C) in addition to the two main subclusters (A3921-A and A3921-B) that are already known. The re-determined mass ratio between the main components A and B is $\sim$2:1. Similar to previous studies of galaxy clusters, we find that a large fraction of the disk galaxies close to the cluster core show no detectable star formation. These are likely systems that are quenched due to ram pressure stripping. Interestingly, we also find quenched spirals at rather large distances of 3-4 Mpc from the cluster core.}
   {A3921-C might be a group of galaxies falling onto the main cluster components. We speculate that the unexpected population of quenched spirals at large clustercentric radii in A3921-A and A3921-B might be an effect of the ongoing cluster merger: shocks in the ICM might give rise to enhanced ram pressure stripping and at least in part be the cause for the quenching of star formation. These quenched spirals might be an intermediate stage in the morphological transformation of field spirals into cluster S0s.}

               
\authorrunning{F. Pranger et al.} 
\titlerunning{The galaxy population of the complex cluster system Abell 3921}

\maketitle
%

\section{Introduction}
Over the past 30 years, observations of galaxies in different environments have confirmed a connection between galaxy properties and the density of the environment; for instance, at present a correlation between galaxy morphology and galaxy number density \citep{dressler80} is well established. Galaxies are found in environments of different density: from 0.2$\rho_{0}$ in voids up to 100$\rho_{0}$ in the inner regions of galaxy clusters \citep{boselli06}, where $\rho_{0}$ is the average field density \citep{geller89}. Since massive galaxy clusters are still growing in the present epoch (e.g. \citealt{donnelly01}), they provide a tool for testing hypotheses on the processes driving the evolution of galaxies. In this context, merging cluster systems are of special interest because they not only represent high-density environments but can also show complex dynamics, increased intra-cluster medium (ICM) turbulence, and shock waves (e.g. \citealt{paul11, roettinger96, markevitch02}).\\

By studying galaxy properties and their evolution with time (and galaxy number density), different interaction processes have been distinguished. Galaxy-galaxy interactions can lead to morphological distortions such as warps, bars, or tidal tails due to close spatial encounters of two (or more) galaxies. They are most efficient at low relative velocities, i.e. in groups or pairs of galaxies or in the outer regions of galaxy clusters (e.g. \citealt{toomre72}). In the latter case, galaxy-galaxy interactions can induce increased star formation rates (e.g. \citealt{porter07}) before the galaxies have been stripped of their gas during the process of falling towards the denser core regions of the cluster (see below).\\
Galaxy harassment \citep{moore96} occurs in galaxy clusters when a galaxy is exposed to tidal forces due to the cluster's gravitational potential and consecutive fly-bys with other cluster members in the core regions. This process can affect many properties of a galaxy within a cluster, including the gas distribution and content, the orbital distribution of stars, and the overall shape.\\    
Furthermore, hydrodynamic interactions between the gaseous component of a galaxy and the hot gas trapped in the cluster's potential have been investigated via observations and simulations, such as ram pressure stripping (e.g. \citealt{abadi99}) or galaxy strangulation (e.g. \citealt{mcgee09}). In the former case the movement of the infalling galaxy relative to the intra cluster medium (ICM) causes ram pressure that can (partially) remove the galaxy's gaseous halo and disk. This process can first enhance (e.g. \citealt{kapferer09}) and eventually quench (e.g. \citealt{quilis00}) star formation, depending on density, velocity and timescale.\\
The gaseous halo represents the hot phase of a galaxy's gas distribution that is gravitationally less bound than the cold gas disk. It is thus more easily affected by ICM interaction. This process can be even more efficient when a galaxy experiences efficient (i.e. low-velocity) tidal interactions with other cluster members in the outer regions of a cluster (e.g. \citealt{balogh00}). After removal of the gaseous halo the only internal gas reservoir is the gas disk which eventually is consumed via star formation - the galaxy gets 'strangulated'. This process is characterised by a gradual decline of the star formation rate (SFR) on relatively long timescales (e.g. \citealt{berrier09}).\\ 

Although at different rates, galaxy-galaxy interactions and ram pressure stripping have been shown to occur also in galaxy subclusters and galaxy groups (e.g. \citealt{fujita04}), thereby pre-processing the galaxies with respect to a future cluster infall. Compared to other environments, (compact) groups of galaxies have been found to show the highest fraction of morphologically distorted galaxies (e.g. \citealt{mendes94}).\\ 
Merger-induced shocks in the intra cluster medium have been observed for more than ten years (e.g. \citealt{markevitch02}). These shocks lead to strong increases in temperature and density in the ICM. It is still a matter of debate whether ICM shocks could increase the efficiency of ram pressure stripping and thereby have an influence on star formation rates and galaxy morphology even in the intermediate density regimes of galaxy cluster infall regions. However, ram pressure stripping has been observed to be efficient even in the outskirts of galaxy clusters up to clustercentric distances of several virial radii (e.g. \citealt{kantharina08}).\\

In this paper we present a spectroscopical and, for the first time, morphological analysis of the galaxies in the merging cluster Abell 3921.\\
Abell 3921 is classified as an Abell cluster of richness 2 and BM-type II at z$\simeq$0.093. Follow-up XMM-Newton/EPIC observations \citep{belsole05} confirmed a bimodal galaxy distribution and, together with numerical simulations \citep{kapferer06}, led to the conclusion that Abell 3921 represents an off-axis merger between a main cluster and a less massive subcluster falling in from the south. Combined optical and X-ray studies have established a mass-ratio of 5:1 \citep{ferrari05}.\\   
Throughout this paper we assume $H_{0}=70$ km/s/Mpc, $\Omega_{m}=0.3$, $\Omega_{\Lambda}=0.7$. At the systemic cluster redshift of 0.093, 1 arcmin corresponds to $\sim$104 kpc in this cosmology. 

\section{The data}
Multifibre spectroscopical service observations in the central 2x2 deg$^{2}$ of A3921 were carried out using the Two Degree Field (2dF) system on the AAT \citep{lewis02}. We used the 400 fibre positioner and double spectrograph system of 2dF to perform two sets of observations both centred at $\alpha=22^{h}49^{m}45.84^{s}$ $\delta = -64^{\circ}22^{m}16.32^{s}$. Fibre allocation was carried out using the 2dF \texttt{Configure} programme, which optimises fibre placement based on user-defined weights associated to each target present in the input catalogue and on a set of instrumental limitations (e.g. minimum distance between allocated fibres). The required input file was created using the catalogue derived from our deep imaging observations (V, R and I passbands, mean seeing: 1.25'', magnitude limit: 22.7 for R band imaging used in this paper; for details see \citealt{ferrari05}). It included a list of fiducial stars with $R_{AB} = 13.7-14.7$, blank sky positions and programme sources, i.e. objects classified as galaxies in our imaging catalogue. The highest weight (i.e. priority) was given to galaxies brighter than $R_{AB} = 18.5$ in order to reach the highest possible completeness level at this magnitude cut ($R^{*}_{AB} +2.1$).\\
We used the same 2dF grating (i.e. 300B) in the two available spectrographs of 2dF and we adopted a central wavelength of 6000 $\mathring{A}$. Our observations thus gave 200 spectra per observing run and spectrograph with a spectral resolution of 8-9 $\mathring{A}$ and covered the approximate wavelength range 3800-8200 $\mathring{A}$ which includes all spectral features from [OII] 3727 $\mathring{A}$ to H$\alpha$ 6563 $\mathring{A}$ at A3921 redshift (z$\approx$0.093). Two runs of observations were carried out in October 2004 and November 2004 with a total exposure time of 3600 seconds per run, divided in three exposures of 1200 seconds in order to eliminate cosmic rays. The data were reduced at the telescope using the 2dF data reduction pipeline software \texttt{2dfdr}.

\subsection{Redshift determination}
We determined redshifts using both cross–correlation against template spectra \citep{tonry79} and identification of possible emission lines. We first used the automatic redshift code for 2dF spectra, \texttt{runz} \citep{colless01}. Then, in order to do a visual check of every best automatic redshift, we measured radial velocities with the IRAF task \texttt{xcsao}\footnote{For emission line galaxies we also used the task \texttt{emcsao}.} of the RVSAO package. In this second step, the results of \texttt{runz} were adopted as initial guess for radial velocities. In both cases, we used as radial velocity standards the eight templates adopted by \citet{colless01}, which include five galaxies and three stars covering a broad range of spectral types. A good agreement between the two different measures was obtained, with a mean offset of 16.9 km/s and a standard deviation of 104.2 km/s for our high quality redshift catalogue (velocity flag (VF) 0 objects - see below). In the following, the measurements derived through the IRAF task \texttt{xcsao} and checked by eye are given and used.\\
A total of 597 redshift estimates have been obtained, among which 546 were derived from high S/N spectra and thus are very precise (VF=0), 24 were obtained from lower S/N spectra, which allowed a redshift measurement but with a higher uncertainty (VF=1), and 27 correspond to bad S/N spectra, not sufficient to get a reliable redshift estimate (VF=2). Among the 597 objects of the total catalogue, 74 are stars (67 with VF=0, 6 with VF=1 and 1 with VF=2), 479 are galaxies with very good z-determination (VF=0), while 18 and 26 are respectively the galaxies with an uncertain (VF=1) or bad (VF=2) z estimate. 61 objects of the catalogue were observed twice, the total number of non-blank sky spectra observed being 658. In all these cases, we kept the measurement characterised by the highest value of $R$, i.e. the parameter related to the S/N of the correlation peak given by \texttt{xcsao}. In summary, 91.6$\%$ of the galaxy redshifts are of high quality, 3.4$\%$ are of low quality and for 5$\%$ it was not possible to get a reliable redshift estimate.\\ 
Among the 61 couples, the mean offset in velocity between the VF=0 objects is -24.8 km/s with a standard deviation of 97.0 km/s. Based also on the previous comparison between \texttt{xcsao} and \texttt{runz} results, we therefore conclude that radial velocities ($cz$) are measured with a typical error of $ \sim $100 km/s. As in \citet{barrena07} the nominal errors given by \texttt{xcsao} show generally lower values, ranging between 10.5 km/s and 121.8 km/s with a mean uncertainty of 48.6 km/s, a factor of $\sim$2 lower than expected. In order to further test this result, we estimated the ratio between a) the difference in the radial velocities measured with \texttt{xcsao} and \texttt{runz} for the VF=0 objects, and b) the corresponding nominal \texttt{xcsao} errors, obtaining a mean value of 1.8. In the following analysis we therefore multiply the nominal \texttt{xcsao} errors by a factor of 2.\\
A3921 has been the object of previous spectroscopical observations \citep{dacosta91,dalton94,katgert96,mathewson96,katgert98,muriel02,ferrari05,way05,pimbblet06}. It is important to stress that all previous observations are however characterised by a lower spectral coverage ($ \sim $3500-6800 $\mathring{A}$).\\
Optical imaging has been taken with the Wide Field Imager camera at the ESO 2.2 m telescope at La Silla observatory. For details see \citet[Sect. 2.1]{ferrari05}.

 \section{Redshift distribution and cluster membership}
\begin{figure}
   \centering
   \includegraphics[angle=270,width=\columnwidth]{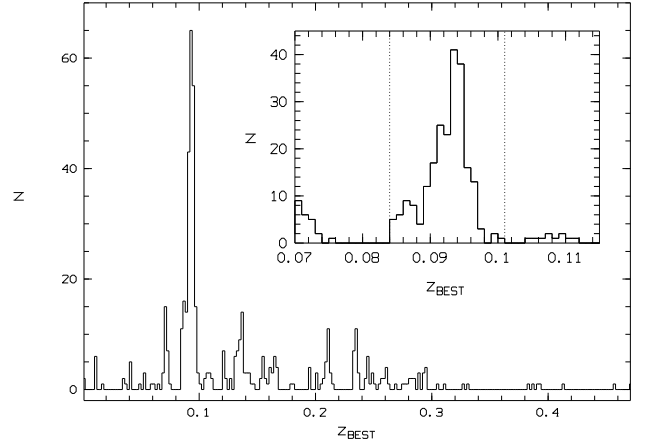}
      \caption{Redshift histogram of the 479 VF=0 galaxies in the 2x2 square degrees field centred on A3921 with a binning of $\Delta$z=0.002. \textit{Inlay:} redshift histogram of the VF=0 galaxies in the redshift range 0.0841$\leq$z$\leq$0.1006 with a binning of $\Delta$z=0.001. The dashed lines indicate the redshift limits for cluster members derived via the the 3$\sigma$ clipping technique (see text for details).}
         \label{fig:479hist}
   \end{figure}

The redshift distribution of the 479 VF=0 galaxies in the 2x2 square degrees field centred on A3921 is shown in Fig. \ref{fig:479hist}. In agreement with previous results \citep{ferrari05,pimbblet06}, the bulk of the cluster is clearly visible as a concentration between z$ \sim $0.082 and z$ \sim $0.102. In order to identify cluster members, we applied the standard iterative 3$\sigma$ clipping \citep{yahil77} as well as a modified 3$\sigma$ clipping involving the biweight estimators for location and scale \citep{beers90} to an initial galaxy sub-sample with redshifts in a wider range (0.075$\leq$z$\leq$0.110). Both versions of the 3$\sigma$ clip reduced the sample from 231 to 223 objects in the redshift range 0.0841$\leq$z$\leq$0.1006. All these objects also satisfy the technique based on gaps in the velocity histogram proposed by \citet{zabludoff93} to filter out foreground and background galaxies. In agreement with the results of previous 2dF observations of A3921 \citep{pimbblet06}, these two simple methods, which make use of only one parameter (the apparent radial velocity of galaxies), indicate as cluster members those galaxies (223) in the redshift range 0.082$\leq$z$\leq$0.102.
    
\subsection{Cluster substructure}
\label{susec:clusub}
\begin{figure}
   \centering
   \includegraphics[angle=0,width=\columnwidth]{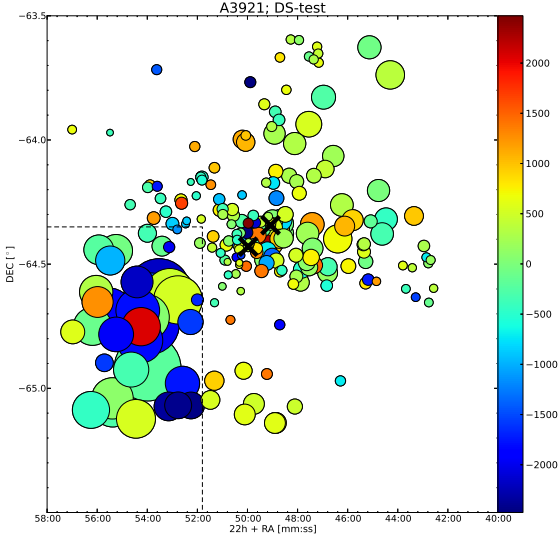}
      \caption{Graphical representation of the DS-test output. The plot shows the spatial location of the galaxies. The radii of the plotted circles are proportional to $e^{\delta}$ where $\delta$ is the DS-test measure of the local deviation from the global velocity dispersion and mean recessional velocity, i.e. larger symbols correspond to a higher significance of residing in a substructure. The colours indicate the rest-frame velocity relative to the cluster centre (BCG1) in km/s. BCG1 and BCG2 are marked by black crosses, respectively. Obviously there is some substructure detected in the south-east region of the cluster (indicated by the dashed lines). For a comparison with subcluster membership see Fig. \ref{fig:cab}.  
              }
         \label{fig:DS}
   \end{figure}

\begin{figure}
   \centering
   \includegraphics[angle=270,width=\columnwidth]{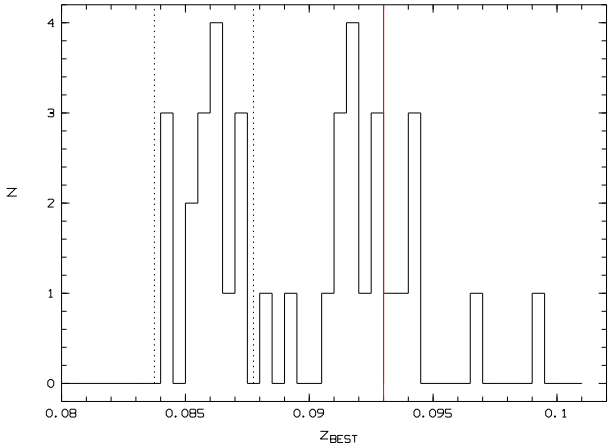}
      \caption{Distribution of the 37 galaxies within the rectangular segment indicated in Fig. \ref{fig:DS} in redshift space. The dashed vertical lines confine the range of objects assigned to component C. The red solid line refers to the median cluster redshift of z=0.093. The two galaxies between z=0.088 and z=0.090 were considered ambiguous members of component C and therefore discarded.
              }
         \label{fig:c_crit}
   \end{figure}

\begin{figure}
   \centering
   \includegraphics[angle=270,width=\columnwidth]{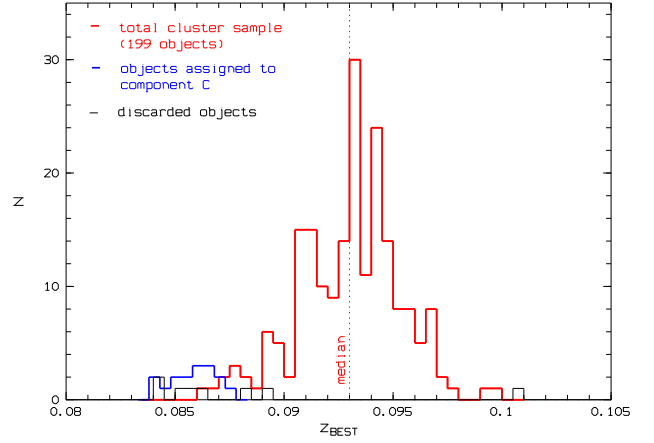}
      \caption{Redshift distribution of the final cluster sample. Red: adjusted total sample of A3921 in z-space. Blue: objects assigned to component C. Black thin line: objects clipped after removal of component C. The two galaxies around z=0.089 are ambiguous members of component C. The median redshift of the adjusted total cluster sample is indicated by the dashed vertical line.
              }
         \label{fig:199_group_discard}
   \end{figure}

Earlier investigations \citep{belsole05,ferrari04,ferrari05,ferrari06} identify A3921 as a merging system that consists of two galaxy clusters (A3921-A, A3921-B; mass ratio $\sim$ 5:1), each containing a giant elliptical galaxy (BCG1, BCG2) coincident with the respective peaks in X-ray emission \citep{belsole05,ferrari04} within $\sim$22 arcseconds (corresponding to $\sim$38 kpc at the cluster redshift) on average. To further investigate the substructure of A3921 on the basis of our new 2dF data we applied the Dressler-Shectman (DS-)test \citep{dressler88} to our sample of 223 confirmed cluster galaxies as a whole. An illustration of this test's outcome is given in Fig. \ref{fig:DS}.\\  
The DS-test reveals some 3D-substructure (significance value P$<5 \cdot 10^{-5}$ for $10^{6}$ Monte-Carlo iterations; this corresponds to$>99.99\%$) in the south-east of the cluster region between $22^{h}51^{m}$ and $22^{h}56^{m}$ in right ascension and $-64^{\circ}24^{m}$ and $-65^{\circ}10^{m}$ in declination. The newly detected clump of galaxies is further referred to as 'component C' or 'A3921-C'. We assigned 16 galaxies to component C using a combined position-redshift space criterion. We first isolated all 37 galaxies within a projected rectangular region confined to the south and east by the borders of our field of view and to the north and west by the lines of longitude at $22^{h}51^{m}48^{s}$ and latitude at $-64^{\circ}21^{m}$, respectively (see Fig. \ref{fig:DS}), thereby keeping all galaxies with $\delta \geq 3$ (see caption of Fig. \ref{fig:DS}). Second, motivated by the redshift distribution shown in Fig. \ref{fig:c_crit}, we introduced a limit of z=0.0875 in redshift space. All 16 galaxies with redshifts lower than 0.0875 were assigned to A3921-C. The two galaxies at z=0.0891 and z=0.0885 were considered ambiguous members of component C according to their intermediate position in z-space and therefore discarded. Reapplying the 3$\sigma$ clippings on the remaining sample of 205 galaxies reduces the number of cluster members to 199. For the separation of this sample into A3921-A and A3921-B we simply measure the projected distance of each galaxy to BCG1 and BCG2, respectively. We do not have any other means for separation due to the similar redshift distributions of components A3921-A and A3921-B. The redshift histogram in Fig. \ref{fig:199_group_discard} illustrates the z-distribution of our final cluster sample as well as the rejected objects or those assigned to component C, respectively. Follow-up DS-tests on the subsamples show significance values of P=0.19 and P=0.59 for A3921-A and A3921-B, respectively, using $10^{6}$ Monte-Carlo iterations. These values are above the generally adopted threshold of P=0.1. The DS-test hence confirms that the redefined cluster samples of A3921-A and A3921-B do not contain substructure.\\

\begin{figure}
   \centering
   \includegraphics[angle=0,width=\columnwidth]{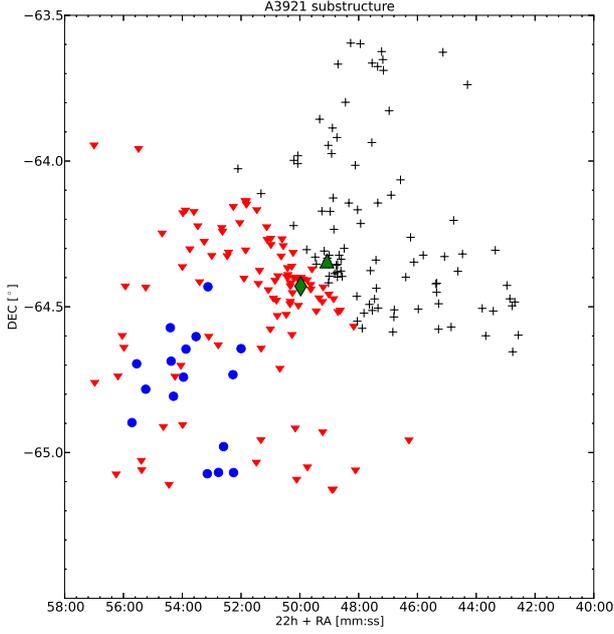}
      \caption{Substructure of A3921. Red triangles indicate objects belonging to subcluster A, black crosses represent galaxies in subcluster B and blue circles show objects assigned to component C. BCG1 and BCG2 are indicated by a green diamond and a green triangle, respectively.
              }
         \label{fig:cab}
   \end{figure}

\begin{figure}
   \centering
   \includegraphics[angle=270,width=\columnwidth]{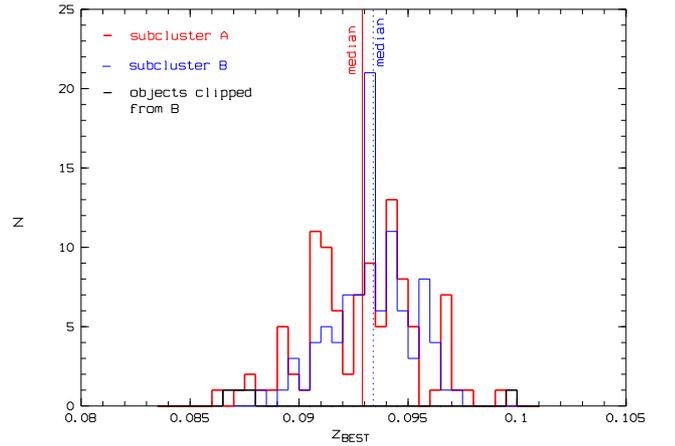}
      \caption{Structure of subclusters A and B in z-space. The black line shows 4 objects clipped from the B-sample. The median redshifts of the samples are indicated by the thin red and the dashed blue line, respectively.
              }
         \label{fig:subAsubBdiscB}
   \end{figure}

Reapplying the 3$\sigma$ clippings to the cluster subsamples reduces the galaxy number in A3921-B by 4 whereas A3921-A is not affected. In the end we have 101 galaxies assigned to A3921-A and 94 galaxies within A3921-B. Fig. \ref{fig:cab} shows the spatial distribution of the subdivided sample, Fig. \ref{fig:subAsubBdiscB} illustrates the structure of A3921-A and A3921-B in redshift-space. Tab. \ref{table:tab1} lists the median redshift and the biweight estimators for scale ($S_{BI}$, corresponding to the velocity dispersion) as well as estimations for radius and total mass calculated on the assumption of isothermal spheres \citep{carlberg97} and on the basis of velocity dispersion and 3D harmonic radius \citep{small98}. The sum of the masses of A3921-A and A3921-B clearly exceeds the mass estimate for the whole cluster. This is due to the estimation technique and reflects the lack of a physical separation criterion for subclusters A and B. Note that the velocity dispersion is slightly lower for the whole sample than for A3921-A in compliance with previous results by \citet{ferrari05}. This surprising result can be explained by the individual redshift distributions of components A and B which have a similar width but are differently skewed. Consequently the estimate for $r_{200}$ is slightly lower for the whole sample than for A3921-A. The respective values are, however, compatible within the error. Also note that our estimates for $r_{200}$ of A3921-A and A3921-B are larger by a factor of $\sim$4 than the results of \citet{ferrari05}. The mass estimates for A3921-A and A3921-B by \citet{ferrari05} are lower than ours by factors of $\sim$4 and $\sim$10, respectively. Note that the mass estimate for A3921-A by \citet{ferrari05} based on velocity dispersion is in agreement with the X-ray mass estimate by \citet{belsole05}. Our estimate for $r_{200}$ of A3921-(A+B) is in excellent agreement with an estimate using the caustics method \citep{diaferio99} on galaxies between 23000 km/s and 33000 km/s in $cz$-space, where $r_{200}$=1.62 Mpc is found. The same technique yields a mass estimate of $1.28 \cdot 10^{14} M_{\odot}$ for A3921-(A+B) which differs from our result. However, note that our computations are based on a virialised system which is probably not fulfilled. Hence, cluster masses might be overestimated in our approach.\\ 
The caustics method (for details see \citealt{diaferio99, serra11}) is an approach to determine cluster membership and estimate mass even for the outskirts of galaxy clusters at distances up to several Mpc from the cluster centre. In such regions the usual assumption of virialisation does not hold and mass estimates relying on this assumption become inaccurate. Caustics are the boundaries of a trumpet-shaped area populated by cluster galaxies in the redshift space diagram (projected clustercentric distance vs. redshift). The caustics method is based only on redshifts and positions. At each distance from the cluster centre it relates the width of the trumpet-shape in redshift to the escape velocity of a given cluster. Masses are estimated by assuming that clusters are spherically symmetric and form by hierarchical clustering.\\ 
As a result of the analysis of our new 2dF data, the mass ratio $R_{M}$ between A3921-A and A3921-B (excluding component C) becomes $R_{M}$=1.90$^{+0.02}_{-0.01}$ whereas \citet{ferrari05} found $R_{M}\sim$5.\\ 
A numerical simulation of the A3921 system based on a mass ratio of $\sim$5 and including galactic winds and star bursts \citep{kapferer06} suggests that the merging process of A3921-A and A3921-B is in a phase where the closest encounter has not happened yet.\\ 
The caustics method has also been used to determine cluster membership based on a galaxy catalogue of A3921 containing 372 galaxies including the objects from our dataset in the $cz$-range between 23000 km/s and 33000 km/s \citep{serra13}. Comparing the resulting sample of 195 galaxies with A3921-(A+B+C) which consists of 215 objects, 119 galaxies were identified as cluster members. Note that all galaxies assigned to A3921-C are among the non-cluster members. The redshift distribution of these 96 non-cluster member objects ranges from z=0.084 to z=0.100 and is characterised by a median value of 0.094 as well as a mean of 0.093 with a standard deviation of 0.003.\\
Since the main science driver of this paper is not the identification of the virialised cluster galaxies, but the impact of the complex environment of A3921 on the galaxy properties, we will in the following use the definition of cluster membership presented earlier in this section.

\begin{table}    
\centering                          
\begin{tabular}{c c c c c c}        
\hline\hline                 
Subsample & $N_{gal}$ & $<z>$ & $S_{BI}$ & $r_{200}$ & $M$ \\&&&[km/s]&[kpc]&[$10^{15} M_{\odot}$]\\    
\hline            
\noalign{\smallskip}   
   Whole sample & $199$ & $0.0932$ & $687^{+51}_{-65}$ & $1627^{+121}_{-154}$ & $2.03^{+0.07}_{-0.06}$\\      
\noalign{\smallskip}  
   A3921-A & $101$ & $0.0929$ & $754^{+66}_{-89}$ & $1786^{+156}_{-211}$ & $1.73^{+0.13}_{-0.12}$\\
\noalign{\smallskip}    
   A3921-B & $94$ & $0.0934$ & $529^{+50}_{-71}$ & $1253^{+118}_{-168}$ & $0.91^{+0.06}_{-0.06}$\\
\noalign{\smallskip} 
 Component C & $16$ & $0.0860$ &  &  & \\
\noalign{\smallskip}    
\hline                                   
\end{tabular}
\caption{Velocity dispersion ($S_{BI}$), $r_{200}$ and $M$ for A3921-A, A3921-B and the total sample. Additionally the number $N_{gal}$ of 3$\sigma$-member galaxies and the median redshift are given for both subclusters, the total sample and component C. We do not attempt to compute $S_{BI}$, $r_{200}$ or $M$ for component C due to the small number of identified members.
 }             
\label{table:tab1} 
\end{table}            
            
\subsection{Field sample definition}
Initially our field sample consists of 256 galaxies left over after deploying the aforementioned limitation procedures on the total galaxy sample, i.e. we neither include the objects supposedly belonging to A3921-C nor objects being ruled out by individual 3$\sigma$ clippings on the subsamples A3921-A and A3921-B. The initial field sample strongly differs from the cluster sample in its redshift- and magnitude distributions. To allow a fair comparison, we choose upper and lower redshift limits on the field sample such that the field and cluster sample have similar luminosity distribution and the same \textit{mean} redshift. Thereby the field sample gets reduced to 83 galaxies. This basic field sample is being used whenever we present comparisons between the field and either the whole cluster (A3921-A + A3921-B) or one of the subclusters, respectively. All further subdivisions applied to the cluster sample analogously apply to the basic field sample. For our morphological analysis the field sample reduces to 74 galaxies due to quality lacks in the respective optical images. This sample will be referred to as the morphological field sample.

\section{Spectroscopical analysis}

\subsection{OII and H$\alpha$ equivalent width measurement}
We determined emission line equivalent widths (EWs) from our 2dF spectra in the following way: The continuum signal was measured within two intervals of 10 $\mathring{A}$ width 20 $\mathring{A}$ left and right from a given emission line, respectively. Next a Gaussian fit was applied to the emission line (in the case of the [OII]-doublet we used a double Gaussian fit with equal FWHM). This fit, as well as the final computation of the EW, was based on the assumption that the continuum intensity at the position of the emission line was the mean of the continuum intensity in the two windows left and right of the emission line.\\
For eight objects with low S/N ratios, the [OII] line could not be fitted with a doublet but only with a single line profile. We corrected the [OII] EWs of these galaxies with a factor of 1.25 determined from 25 objects with the highest S/N, for which we compared robust doublet fits to singlet fits.\\
We estimated the EW upper limits of [OII] and H$\alpha$ by simulating spectra with lines of a FWHM of 4.3 $\mathring{A}$ (our dispersion) and continuum fluxes and noise from spectra for which we obtained the lowest EW values. This leads to two different estimates for [OII] and H$\alpha$. We simulated EWs between 0.25 $\mathring{A}$ and 25 $\mathring{A}$. Requiring that input and output EW had to agree within the measurement errors, we thus found a minimum EW of 4.9 $\mathring{A}$ for [OII] and 1.5 $\mathring{A}$ for H$\alpha$.      

\subsection{Equivalent widths and star formation rates}
\label{susec:radial}
\begin{figure}
   \centerline{\includegraphics[angle=270,width=\columnwidth]{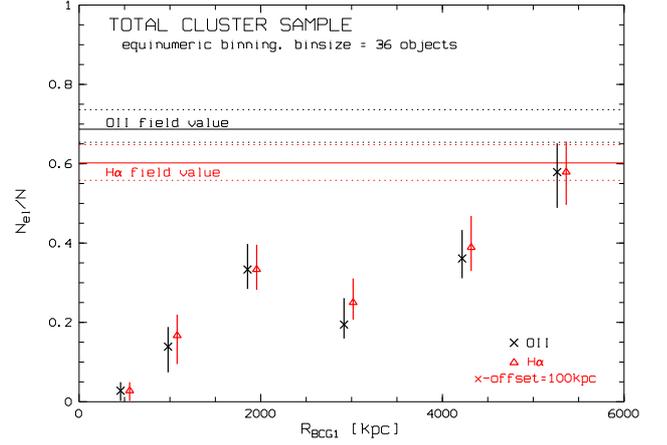}}
      \caption{Fraction of galaxies showing [OII] or H$\alpha$ emission lines, respectively, vs. distance from the cluster centre. The corresponding data for the subsamples of components A and B are listed in Tab. \ref{table:tab2}.   
              }
         \label{fig:frac_tot}
   \end{figure}

\begin{figure}
   \centerline{\includegraphics[angle=270,width=\columnwidth]{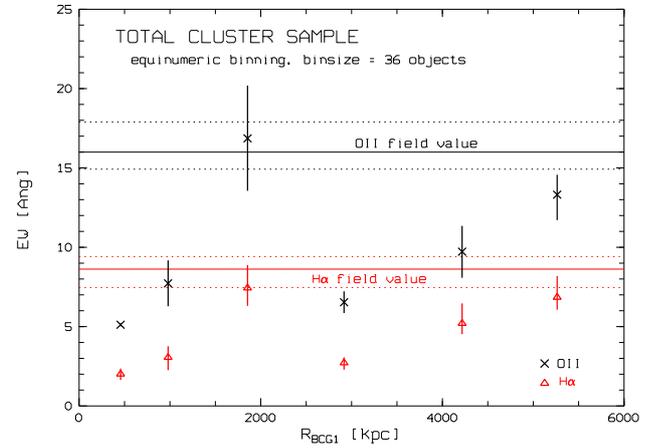}}
      \caption{Mean equivalent width of [OII] and H$\alpha$ emission lines for all objects (objects without detectable lines get assigned a lower threshold value). The corresponding data for the subsamples of components A and B are listed in Tab. \ref{table:tab2}.}
         \label{fig:ew_tot}
   \end{figure}

\begin{figure}
   \centerline{\includegraphics[angle=270,width=\columnwidth]{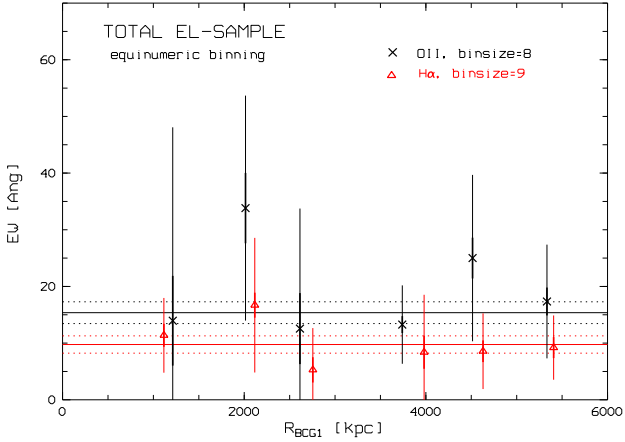}}
      \caption{Median equivalent width of [OII] and H$\alpha$ emission lines only for EL-objects. The thin error bars represent the standard deviation within each data bin. The corresponding data for the subsamples of components A and B are listed in Tab. \ref{table:tab3}. The horizontal lines represent the field values and 1$\sigma$ errors.    
              }
         \label{fig:oew_tot}
 \end{figure}

\begin{figure}
   \centerline{\includegraphics[angle=270,width=\columnwidth]{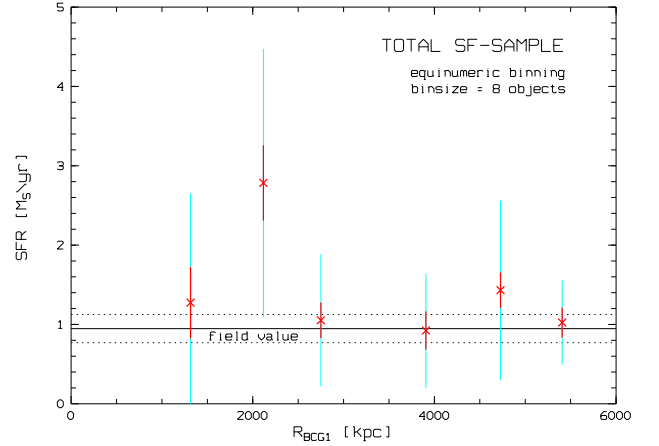}}
      \caption{Median star formation rate for objects with detectable [OII] emission line. The thin error bars represent the standard deviation within each data bin. The corresponding data for the subsamples of components A and B are listed in Tab. \ref{table:tab3}.
              }
         \label{fig:sfr_tot}
 \end{figure}

Figs. \ref{fig:frac_tot} to \ref{fig:sfr_tot} show the projected radial trends (with respect to the centre of A3921-A and of the whole sample taken to be coincident with BCG1) of 
\begin{enumerate}[a)]
 \item the fraction of EL-galaxies (i.e. galaxies definitively showing [OII], H$\alpha$ or both emission lines) per bin,
 \item the equivalent width of [OII] and H$\alpha$ lines and 
 \item $[$OII]-based star formation estimates 
\end{enumerate}
for the total sample (i.e. A3921-(A+B)). The corresponding data for both subsamples are given in Tabs. \ref{table:tab2} and \ref{table:tab3}. To estimate the star formation rates (SFR) of galaxies we adopt a relation presented in \citet{kennicutt92}:

\begin{equation}
      $SFR$(M_{\odot}yr^{-1})\simeq 2.7 \cdot 10^{-12} \frac{L_{R}}{L_{R}(\odot)} EW([OII]) E(H\alpha)
      \label{eq:sfr}
\end{equation}

where $EW([OII])$ denotes the z-corrected equivalent width of [OII] and $E(H\alpha)$ stands for the extinction value at the respective wavelength. To apply this measure to our data we first had to determine the R-band absolute magnitudes $L_{R}$. For this purpose we used the image processing software \texttt{SExtractor} \citep{bertin96} calibrated via magnitudes of 37 objects we could identify in the magnitude catalogue appended to \citet{ferrari05}. Moreover, for a proper k-correction all spectra were classified by comparison with spectra from \citet{kennicutt92b}.\\ 
All errors given in Figs. \ref{fig:frac_tot} to \ref{fig:sfr_tot} and Figs. \ref{fig:TOT_c} to \ref{fig:TOT_m20} as well as in Tabs. \ref{table:tab2}, \ref{table:tab3} and \ref{table:tab4} are generated via bootstrapping techniques except for bins with less than ten objects where the standard deviation error is additionally given (indicated by thin error-lines in the plots). All clustercentric distances ($R_{BCG1}$) are median values for each bin. To rule out potential binning biases we analysed different binning methods (equidistant and equinumeric binning) and a wide range of binsizes. The plots shown here reflect trends that are robust against binning changes. We use an equinumeric binning in these plots, i.e. whenever we show a given quantity as a function of radius, we keep the number of galaxies per bin fixed throughout the plot, except for the last data point which contains the respective remainder according to the total number of objects within the sample under investigation. Horizontal lines indicate the respective field values. Fig. \ref{fig:frac_tot} illustrates fractions of galaxies showing emission lines. For the first EW-plot (i.e. Fig. \ref{fig:ew_tot}), we assigned an equivalent width of 5.1 $\mathring{A}$ to objects without a detectable [OII] doublet, and an EW of 1.6 $\mathring{A}$ to the objects without a detectable H$\alpha$ line. These values correspond to the lowest measured EWs of [OII] and H$\alpha$, respectively, and are justified by the mock spectra discussed at the end of the last section. For the second EW-plot (i.e. Fig. \ref{fig:oew_tot}) only objects with detectable [OII] or H$\alpha$ emission lines (constituting the EL-sample) were considered. For the SFR-plots (i.e. Fig. \ref{fig:sfr_tot}) only objects with detected [OII] emission (constituting the SF-sample) were considered.\\

\begin{table*}    
\centering                          
\begin{tabular}{ccccc|ccccc}        
\hline\hline                 
\multicolumn{5}{c|}{Subcluster A} & \multicolumn{5}{c}{Subcluster B} \\
\hline
&&&&&\\[-1.5ex]
$R_{BCG1}$ & $ N_{[OII]}/N $ & $ N_{H\alpha}/N $ & $ EW([OII]) $ & $ EW(H\alpha) $ & $R_{BCG1}$ & $ N_{[OII]}/N $ & $ N_{H\alpha}/N $ & $ EW([OII]) $ & $ EW(H\alpha) $ \\$ $[kpc]&&&[Ang]&[Ang]&[kpc]&&&[Ang]&[Ang]\\ 
\hline           
&&&&&\\[-1ex] 
   $261.6$ & $0$ & $0$ & $5.0$ & $1.6$ & $881.5$ & $0.100^{+0.025}_{-0.035}$ & $0.150^{+0.055}_{-0.085}$ & $8.9^{+1.7}_{-2.7}$ & $2.0^{+0.2}_{-0.2}$ \\[1ex]
   $713.2$ & $0.100^{+0.020}_{-0.030}$ & $0.100^{+0.054}_{-0.042}$ & $5.6^{+0.5}_{-0.3}$ & $3.2^{+1.1}_{-0.9}$ & $1725.3$ & $0.400^{+0.055}_{-0.085}$ & $0.400^{+0.087}_{-0.073}$ & $12.7^{+3.6}_{-2.6}$ & $7.6^{+0.9}_{-1.7}$ \\[1ex] 
   $1497.3$ & $0.250^{+0.096}_{-0.096}$ & $0.250^{+0.068}_{-0.040}$ & $18.2^{+6.2}_{-5.2}$ & $6.5^{+1.3}_{-2.5}$ & $2788.9$ & $0.200^{+0.076}_{-0.040}$ & $0.200^{+0.096}_{-0.078}$ & $5.5^{+0.3}_{-0.3}$ & $2.5^{+0.5}_{-0.3}$ \\[1ex] 
   $2719.9$ & $0.200^{+0.050}_{-0.050}$ & $0.250^{+0.110}_{-0.072}$ & $8.3^{+1.8}_{-1.6}$ & $3.7^{+1.3}_{-0.9}$ & $3730.4$ & $0.300^{+0.088}_{-0.064}$ & $0.350^{+0.115}_{-0.083}$ & $9.2^{+2.5}_{-1.1}$ & $4.9^{+1.6}_{-1.2}$ \\[1ex]
   $4339.5$ & $0.550^{+0.052}_{-0.066}$ & $0.600^{+0.069}_{-0.077}$ & $13.3^{+2.3}_{-2.9}$ & $7.2^{+0.9}_{-1.7}$ & $5101.7$ & $0.429^{+0.129}_{-0.107}$ & $0.429^{+0.209}_{-0.211}$ & $10.1^{+0.8}_{-1.8}$ & $5.0^{+1.5}_{-1.5}$ \\[1ex] 
\hline
\hline
\end{tabular}
\caption{Mean equivalent widths and fractions of EL-galaxies (based on [OII] and H$\alpha$, respectively) for subclusters A and B with a binsize of 20 objects. Field values (based on the whole field of view) are shown in Figs. \ref{fig:frac_tot} and \ref{fig:ew_tot}.
 }            
\label{table:tab2} 
\end{table*}

It is noticeable that within a distance of about $ r_{200} $ (for the case of A3921-B it is $r_{200}$+$d$ where $d\simeq$ 850 kpc represents the projected distance between the BCGs) from BCG1 all investigated quantities show an increase towards larger clustercentric radii, in compliance with previous studies on the dependence of galaxy properties on environment (e.g. \citealt{verdugo08}). Furthermore, at a clustercentric distance of $\sim$3 Mpc, fractions of EL-galaxies as well as equivalent widths (see. Figs. \ref{fig:frac_tot} to  \ref{fig:ew_tot} and Tab. \ref{table:tab2}) show a noticeable decline. Also the equivalent widths of EL-galaxies only and the SFR plots (see Figs. \ref{fig:oew_tot} to  \ref{fig:sfr_tot} and Tab. \ref{table:tab3}) show a decrease at these radii, albeit with worse number statistics.\\ 
Comparing both subclusters (especially at radial distances $>$4 Mpc), the investigated quantities reach lower values in A3921-B than in A3921-A, albeit at low statistical significance. However, due to the lack of a physical separation criterion between A3921-A and A3921-B and due to the low number statistics for SFR plots and the second series of emission line plots we do not attempt to further interpret this.\\ 
  
  \begin{table*}    
\centering                          
\begin{tabular}{ccccc|ccc}        
\hline\hline                 
\multicolumn{8}{c}{Subcluster A} \\ 
\hline
&&&&&&&\\[-1.5ex]
$R_{BCG1}$ & $ EW([OII]) $ & $ RMS_{EW(OII)} $ & SFR & $ RMS_{(SFR)} $ & $R_{BCG1}$ & $ EW(H\alpha) $ & $ RMS_{EW(H\alpha)} $ \\$ $[kpc]&[Ang]&[Ang]&[M$\odot$/yr]&[M$\odot$/yr]&[kpc]&[Ang]&[Ang]\\ 
\hline           
&&&&&&&\\[-1ex] 
   $1204.0$ & $17.1\pm1.6$ & $44.6$ & $2.9\pm0.4$ & $1.4$ & $ 1204.0 $ & $12.5\pm1.1$ & $5.8$ \\[1ex]
   $2130.7$ & $48.7\pm3.2$ & $16.7$ & $3.4\pm0.6$ & $1.6$ & $ 2130.7 $ & $27.3\pm0.8$ & $11.6$ \\[1ex] 
   $3035.5$ & $16.6\pm1.8$ & $8.9$ & $1.6\pm0.3$ & $0.8$ & $ 2719.9 $ & $3.1\pm1.7$ & $6.8$ \\[1ex] 
   $4208.6$ & $15.0\pm1.9$ & $10.3$ & $1.3\pm0.3$ & $0.9$ & $ 4037.2 $ & $6.7\pm0.7$ & $4.0$ \\[1ex]
   $4896.0$ & $24.3\pm3.6$ & $17.0$ & $1.5\pm0.2$ & $0.7$ & $ 4322.4 $ & $15.2\pm1.3$ & $4.9$ \\[1ex]
   $5540.7$ & $25.3\pm10.2$ & $14.3$ & $0.9\pm0.1$ & $0.8$ & $ 5333.3 $ & $13.1\pm0.3$ & $4.2$ \\[1ex]
   \hline
\multicolumn{8}{c}{Subcluster B} \\ 
\hline
&&&&&&&\\[-1ex]   
   $1314.8$ & $12.8\pm3.6$ & $9.1$ & $0.9\pm0.6$ & $1.7$ & $ 1115.7 $ & $6.0\pm1.9$ & $7.6$ \\[1ex]
   $1930.3$ & $20.2\pm4.3$ & $13.8$ & $1.4\pm0.2$ & $0.4$ & $ 1744.5 $ & $12.7\pm0.3$ & $7.8$ \\[1ex] 
   $2679.4$ & $9.9\pm9.2$ & $30.5$ & $0.8\pm0.3$ & $0.8$ & $ 2816.1 $ & $6.4\pm3.6$ & $8.2$ \\[1ex] 
   $3562.7$ & $7.8\pm1.1$ & $3.6$ & $0.7\pm0.1$ & $0.3$ & $ 3648.4 $ & $8.8\pm0.8$ & $12.6$ \\[1ex]
   $4515.4$ & $25.0\pm4.4$ & $12.6$ & $1.4\pm0.2$ & $0.5$ & $ 5101.0 $ & $4.4\pm3.0$ & $7.3$ \\[1ex]
   $5258.6$ & $17.7\pm2.3$ & $5.6$ & $1.1\pm0.5$ & $1.5$ & $ 5869.9 $ & $6.9\pm3.6$ & $7.7$ \\[1ex] 
\hline
\hline
\end{tabular}
\caption{Median eqivalent widths and star formation rates in subclusters A and B. Each radial bin contains 4 galaxies (5 for $EW_{H\alpha}$, subcluster B). Field values (based on the whole field of view) are shown in Figs. \ref{fig:oew_tot} and \ref{fig:sfr_tot}.
 }            
\label{table:tab3} 
\end{table*}
 
\section{Morphological analysis}
\label{susec:morph} 

    \begin{figure}
   \centering
   \includegraphics[angle=270,width=\columnwidth]{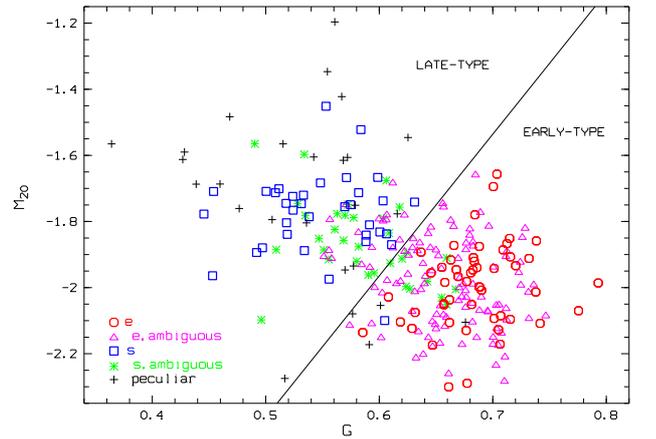}
      \caption{Overall morphological sample (263 objects) in Gini-$M_{20}$ space, morphologically subdivided into spirals (s), ellipticals (e) and peculiars. The black line represents the cut applied to the unambiguously classified galaxies (blue squares and red circles) finally dividing the whole sample into late-type (upper left corner) and early-type galaxies (lower right corner). Objects classified as peculiar are included in the respective subsample. Note that we use this quantitative criterion throughout our morphological analysis.
      }
         \label{fig:g_2m20_tot}
 \end{figure}

       \begin{figure}
   \centering
   \includegraphics[angle=0,width=\columnwidth]{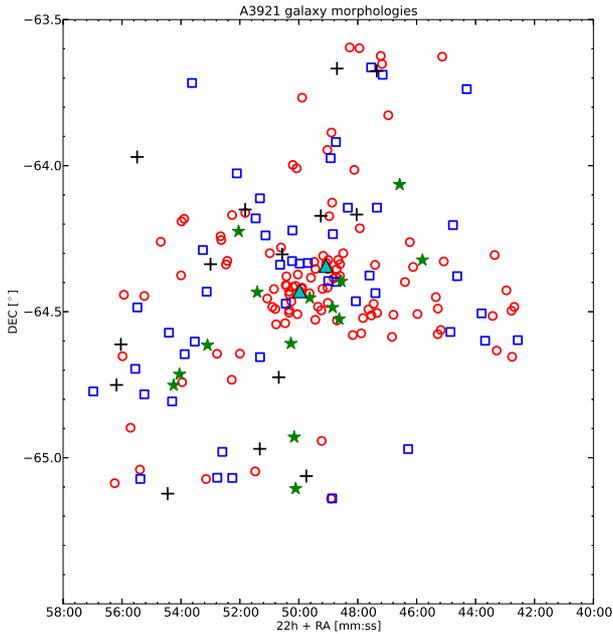}
      \caption{Illustration of galaxy positions and morphological galaxy types. Early-type galaxies are shown as red circles, late-type objects are illustrated as blue squares. No-EL spirals are shown separately as green stars. Black crosses indicate morphologically distorted (i.e. peculiar) galaxies. BCG1 and BCG2 are represented by cyan triangles.   
           }
         \label{fig:visual}
   \end{figure}
 
\begin{figure}
   \centerline{\includegraphics[angle=270,width=\columnwidth]{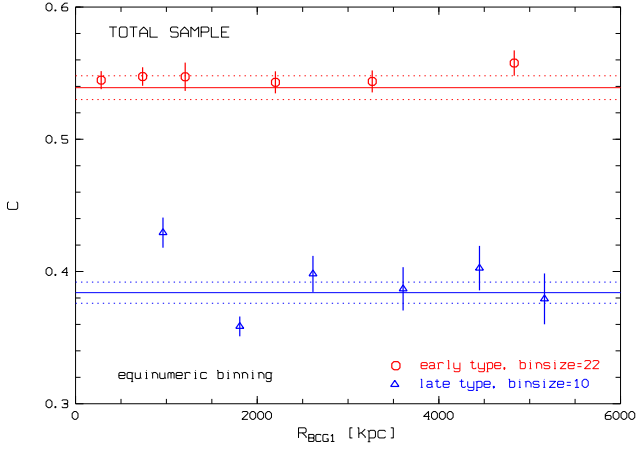}}
      \caption{Concentration index C as a function of clustercentric distance for early-type galaxies (red circles) and late-type galaxies (blue triangles). The corresponding data for the subsamples of components A and B are listed in Tab. \ref{table:tab4}. The horizontal lines represent the field values and 1$\sigma$ errors. 
      }
         \label{fig:TOT_c}
 \end{figure}

\begin{figure}
   \centerline{\includegraphics[angle=270,width=\columnwidth]{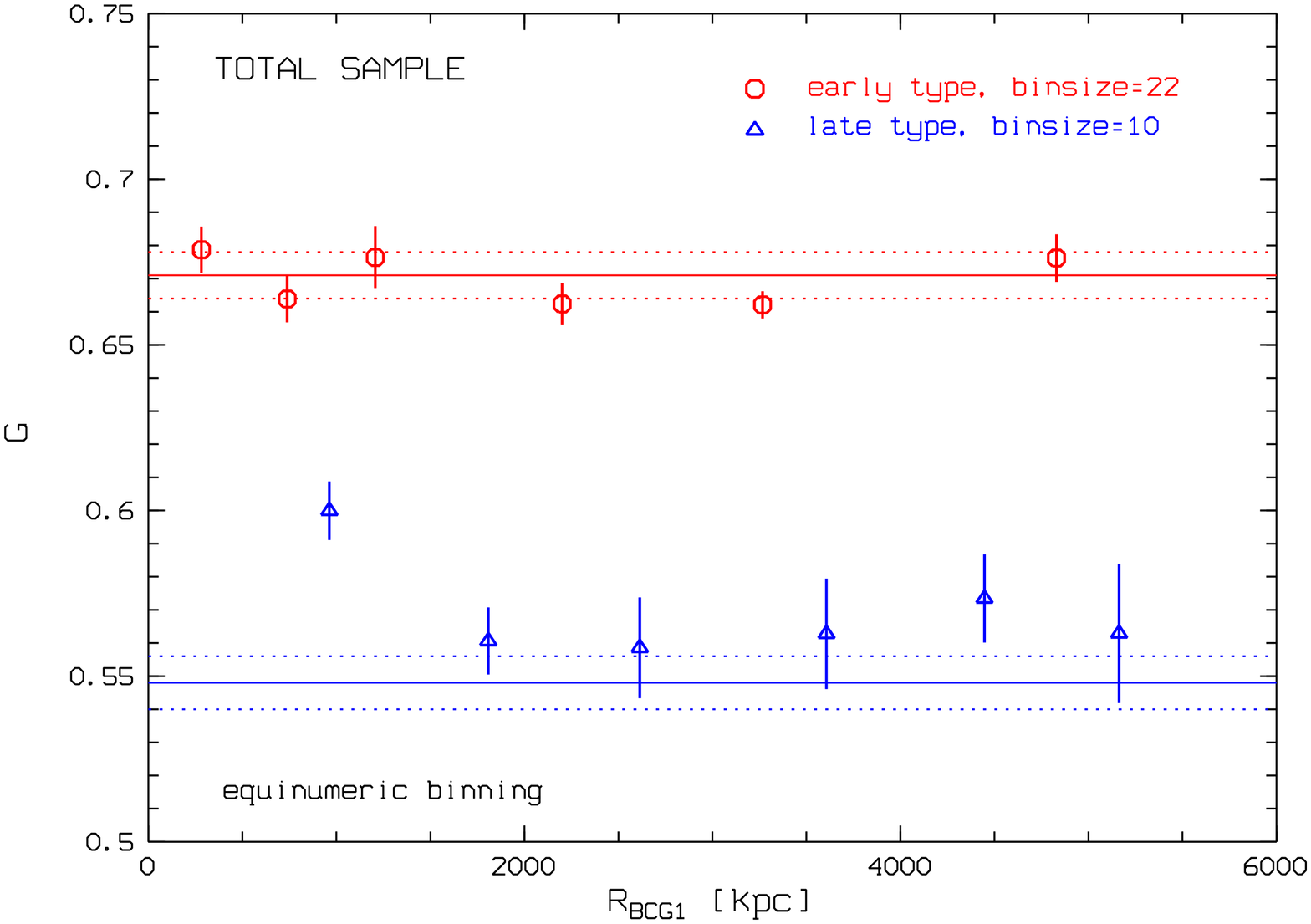}}
      \caption{Gini coefficient G as a function of clustercentric distance for early-type galaxies (red circles) and late-type galaxies (blue triangles). The corresponding data for the subsamples of components A and B are listed in Tab. \ref{table:tab4}. The horizontal lines represent the field values and 1$\sigma$ errors.}
         \label{fig:TOT_g_2}
 \end{figure}

\begin{figure}
   \centerline{\includegraphics[angle=270,width=\columnwidth]{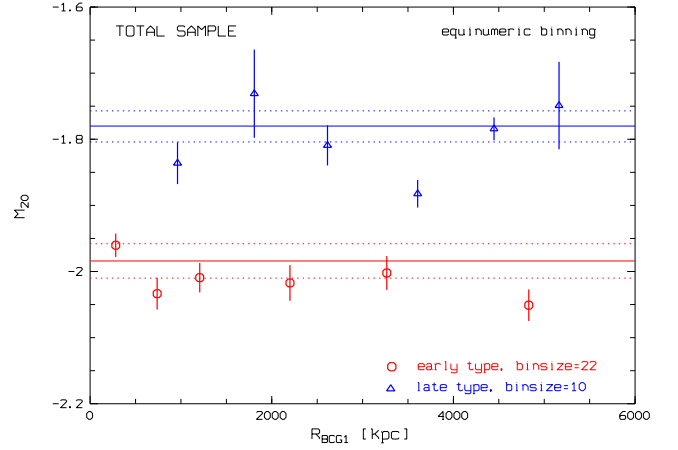}}
      \caption{Second order moment of the brightest 20\% of the galaxy, $M_{20}$, as a function of clustercentric distance for early-type galaxies (red circles) and late-type galaxies (blue triangles). The corresponding data for the subsamples of components A and B are listed in Tab. \ref{table:tab4}. The horizontal lines represent the field values and 1$\sigma$ errors.
           }
         \label{fig:TOT_m20}
   \end{figure}

For our morphological analysis we use the WFI imaging data by \citet{ferrari05}. From the combined sample of cluster and field galaxies (199+83=282 objects) we discard 10 cluster objects and 9 field objects because of corrupted imaging and therefore end up with a total morphological sample of 263 spectroscopically confirmed galaxies. For each of these we calculate Gini coefficient, concentration index and $M_{20}$ index \citep{lotz04} (the resolution of the WFI imaging is inefficient for an estimation of the asymmetry index, we therefore do not use this descriptor in the following). The Gini coefficient (G) was originally introduced to assess the distribution of wealth within a human population - here we use it to describe the distribution of pixel values in a galaxy's image \citep{abraham03}. It relates the Lorenz curve \citep{lorenz1905} of a uniform flux distribution to the observed flux distribution. A Gini coefficient of zero corresponds to a completely uniform distribution while for a Gini coefficient of one, all the flux is contained within a single pixel. The concentration index (C) measures the ratio of flux within an inner and an outer aperture. The $M_{20}$ index ($M_{20}$) is defined as the second-order moment of the brightest 20 percent of the pixels in a galaxy image. This quantity hence is sensitive to bright off-centre features like star-forming regions.\\  
As a start, we visually classify each galaxy as spiral, elliptical or peculiar (i.e. showing morphological distortions). Fig. \ref{fig:g_2m20_tot} shows the morphological sample (cluster and field) in Gini-$M_{20}$ space. Using only cases of unambiguous visual classifications (1/3 of our sample), we apply a dividing line in Gini-$M_{20}$ space to separate between early- and late-type galaxies. This \textit{purely quantitative} morphological classification will be used in the following. In Fig. \ref{fig:visual} the positions and morphological types of all 189 members of the morphological cluster sample are depicted. Figs. \ref{fig:TOT_c} to \ref{fig:TOT_m20} illustrate the radial trends of the morphological quantities for the whole cluster sample. The corresponding data for components A and B are given in Tab. \ref{table:tab4} (as before, the small number statistics do not allow component C to be investigated separately). In the latter case as far as A3921-B is concerned the respective morphological subsample is reduced by 4 galaxies that were excluded from A3921-B by 3$\sigma$ clipping.\\ 
For the analyses of local surface number densities (see Sec. \ref{sec:dens}) we will at first add the objects assigned to component C to the morphological cluster sample and then make use of three different subsamples of this extended morphological sample (EMS): a spiral sample, a no-EL spiral sample and a peculiar sample. The spiral sample consists of all objects unambiguously classified as spirals as well as unclear cases located in the late-type regime of the Gini-$M_{20}$ space. We also define a no-EL spiral sample that holds morphologically classified spirals (i.e. galaxies identified as late-type either unambiguously by-eye or on the basis of their Gini/$M_{20}$ indices) showing neither [OII] nor H$\alpha$ emission line features. Note that the location in the calibrated Gini-$M_{20}$ space is a purely mathematical criterion which we use to classify all members of the morphological sample (field and cluster objects) either as a late-type or as an early-type galaxy. The peculiar sample consists of all galaxies visually identified as peculiar, i.e. showing morphological distortions (see Sec. \ref{susec:noel}). Objects with unambiguous deviations from an undisturbed morphology are classified as peculiar. However, as illustrated in Fig. \ref{fig:g_2m20_tot} the peculiar sample shows a wide spread in Gini-$M_{20}$ space. It is hence impossible to define a quantitative Gini-$M_{20}$ criterion for peculiarity. This is due to the limited spatial resolution of our ground-based imaging. Thus, in contrast to the distinction into late-type and early-type galaxies, we keep a visual classification for the peculiars \textit{only}. \\

  \begin{table*}    
\centering                          
\begin{tabular}{cccc|cccc}        
\hline\hline                 
\multicolumn{4}{c|}{Subcluster A} & \multicolumn{4}{c}{Subcluster B} \\
\hline
\multicolumn{8}{c}{Late-type galaxies} \\
\hline
&&&&&&&\\[-1.5ex]
$R_{BCG1}$ &  C  &  G  & $ M_{20} $ & $R_{BCG1}$ &  C  &  G  & $ M_{20} $ \\$ $[kpc]&&&&[kpc]&&&\\ 
\hline           
&&&&&&&\\[-1ex] 
    $934.2$ & $0.437\pm0.015$ & $0.600\pm0.016$ & $-1.854\pm0.028$ & $ 942.4$ & $ 0.427\pm0.008 $ & $0.609\pm0.006$ & $-1.835\pm0.010$ \\[1ex]
   $1891.5$ & $0.376\pm0.010$ & $0.571\pm0.022$ & $-1.763\pm0.027$ & $1646.8$ & $ 0.351\pm0.018 $ & $0.536\pm0.004$ & $-1.713\pm0.016$ \\[1ex] 
   $2407.0$ & $0.380\pm0.028$ & $0.553\pm0.028$ & $-1.813\pm0.055$ & $2202.4$ & $ 0.340\pm0.003 $ & $0.554\pm0.005$ & $-1.746\pm0.080$ \\[1ex] 
   $3388.3$ & $0.370\pm0.011$ & $0.535\pm0.012$ & $-1.855\pm0.052$ & $3320.3$ & $ 0.415\pm0.003 $ & $0.556\pm0.007$ & $-1.861\pm0.005$ \\[1ex]
   $4128.6$ & $0.375\pm0.019$ & $0.552\pm0.013$ & $-1.830\pm0.041$ & $4375.7$ & $ 0.397\pm0.010 $ & $0.577\pm0.016$ & $-1.788\pm0.016$ \\[1ex]
   $4382.5$ & $0.381\pm0.018$ & $0.574\pm0.014$ & $-1.768\pm0.009$ & $5102.4$ & $ 0.401\pm0.017 $ & $0.571\pm0.013$ & $-1.856\pm0.006$ \\[1ex]
   \hline
\multicolumn{8}{c}{Early-type galaxies} \\
\hline
&&&&&&&\\[-1ex]   
    $156.7$ & $0.571\pm0.009$ & $0.684\pm0.006$ & $-1.942\pm0.010$ & $ 814.9$ & $ 0.592\pm0.006 $ & $0.712\pm0.013$ & $-2.014\pm0.036$ \\[1ex]
   $405.4$ & $0.522\pm0.006$ & $0.649\pm0.006$ & $-1.978\pm0.016$ & $1501.7$ & $ 0.552\pm0.011 $ & $0.688\pm0.009$ & $-1.998\pm0.022$ \\[1ex] 
   $712.7$ & $0.537\pm0.007$ & $0.657\pm0.004$ & $-2.037\pm0.021$ & $2212.1$ & $ 0.546\pm0.019 $ & $0.665\pm0.018$ & $-2.006\pm0.019$ \\[1ex] 
   $1069.2$ & $0.546\pm0.005$ & $0.674\pm0.004$ & $-2.005\pm0.013$ & $2916.4$ & $ 0.543\pm0.013 $ & $0.663\pm0.008$ & $-2.001\pm0.025$ \\[1ex]
   $2188.9$ & $0.544\pm0.007$ & $0.659\pm0.004$ & $-1.992\pm0.025$ & $3631.6$ & $ 0.545\pm0.013 $ & $0.688\pm0.006$ & $-2.008\pm0.041$ \\[1ex]
   $4279.8$ & $0.548\pm0.010$ & $0.651\pm0.005$ & $-2.081\pm0.008$ & $5078.4$ & $ 0.563\pm0.014 $ & $0.676\pm0.013$ & $-2.028\pm0.015$ \\[1ex]
   \hline
\hline
\end{tabular}
\caption{Concentration index (C), Gini coefficient (G) and the second order moment of the brightest 20\% of the galaxy ($M_{20}$ index) for subclusters A and B (median values). \textit{Upper panel:} late-type galaxies (subcluster A: binsize = 4 galaxies, subcluster B: binsize = 5 galaxies). \textit{Lower panel:} early-type galaxies (binsize = 11 galaxies). Field values (based on the whole field of view) are shown in Figs. \ref{fig:TOT_c}, \ref{fig:TOT_g_2} and \ref{fig:TOT_m20}.
 }            
\label{table:tab4} 
\end{table*}  
   
The composition of the A3921-(A+B+C) sample objects identified as non-cluster members by the caustics method (see Sec. \ref{susec:clusub}) with respect to the morphological classification is as follows: 54$\%$ ellipticals, 32$\%$ spirals, 14$\%$ peculiars (i.e. galaxies with morphological distortions), compared to a mix of 74$\%$, 25$\%$ and  1$\%$ in A3921-(A+B+C), respectively. The additional constraints implied by the caustic method would hence reduce our cluster sample by 96 galaxies showing a cluster-like distribution in z-space. The type mix of our sample would thus be significantly changed, in the sense that the fractions of spiral and peculiar galaxies would be reduced. Consequently, the overdensities of spirals and peculiars depicted in Figs. \ref{fig:spir-tot_min}, \ref{fig:pec-tot_min} and \ref{fig:pec-tot_std__min} are not found when we restrict our analysis to the 119 cluster members identified using the caustics method. However, the radial trends presented in Sec. \ref{susec:radial} are still found when analysing the sample of 119 cluster members, albeit with lower significance. Note that only one of the non-cluster member objects is a spiral galaxy without detectable [OII] and H$\alpha$ emission lines (i.e. a no-EL spiral, see Sec. \ref{susec:noel}).\\ 

\subsection{No-EL spirals}
\label{susec:noel}
\begin{figure}
   \centering
   \includegraphics[angle=270,width=\columnwidth]{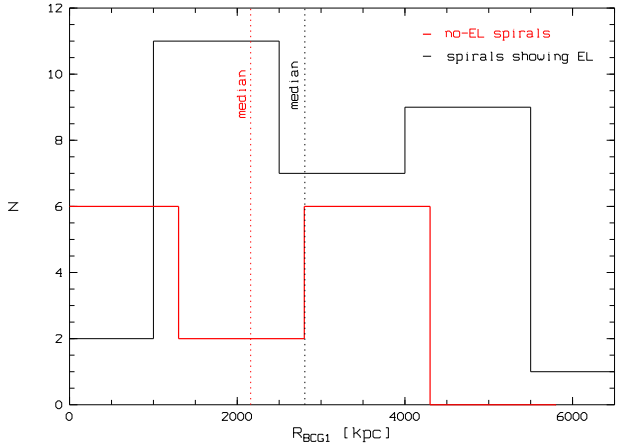}
      \caption{Number of morphologically classified spiral galaxies as a function of clustercentric distance. The sample showing no emission lines is found to be on average closer to the cluster centre. Note that component C is included in this plot.
           }
         \label{fig:el_noel}
   \end{figure}

\begin{figure}
   \centering
   \includegraphics[angle=270,width=\columnwidth]{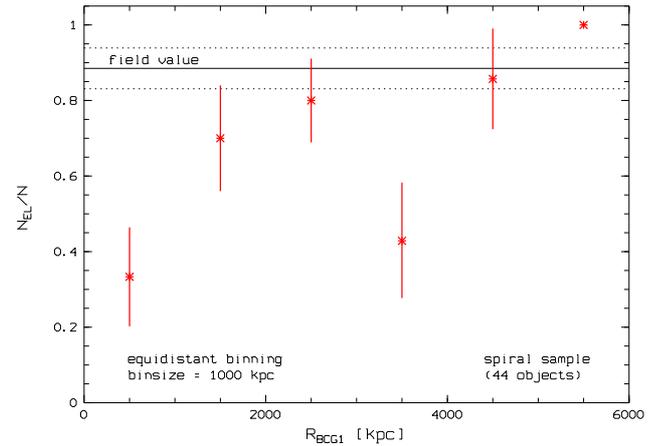}
      \caption{Fraction of morphologically identified spirals showing [OII] or H$\alpha$ emission lines, respectively. Note that component C is included in this plot. Data bins are equidistant. 
           }
         \label{fig:spiral_frac}
   \end{figure}

On average the galaxies morphologically classified as spirals with no detectable [OII] or H$\alpha$ emission are found to be closer to the cluster centre. This trend is depicted in Figs. \ref{fig:el_noel} and \ref{fig:spiral_frac}. At a distance of around 3.5 Mpc we find that the local decrease in Fig. \ref{fig:spiral_frac} (with a corresponding increase in no-EL spiral frequency in Fig. \ref{fig:el_noel}) coincides with similar trends in fractions of EL-galaxies, equivalent width, star formation rate and, most prominent, in $M_{20}$, both in A3921-A and A3921-B. We now take a look at local projected surface number densities to determine the morphological galaxy species that come into question to be the cause for this local decrease.\\

\section{Surface number densities}
\label{sec:dens}
\begin{figure}
   \centering
   \includegraphics[angle=0,width=\columnwidth]{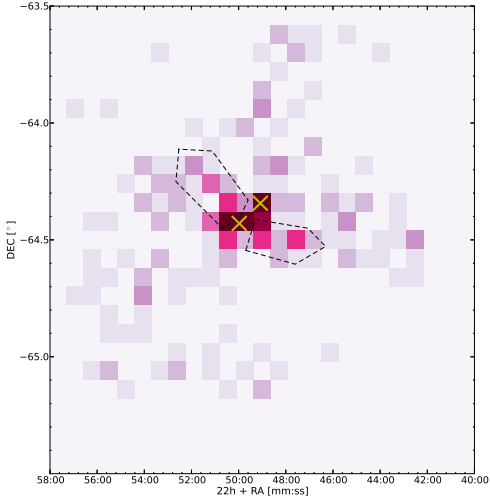}
      \caption{Galaxy number density of the whole spectroscopic sample of A3921 (including component C). The field-of-view is subdivided into 25x25 equally-sized rectangles. The colour scale shows the number of galaxies per Mpc$^{2}$. BCG1 and BCG2 are represented by yellow crosses. Overdense regions protruding from the cluster centre are enclosed by dashed lines.  
      }
         \label{fig:mem_grid_25}
   \end{figure}

Fig. \ref{fig:mem_grid_25} illustrates the projected absolute surface number density of the whole cluster sample including component C (215 galaxies, A3921-(A+B+C)) based on a subdivision of the 2dF field-of-view into 625 equally-sized cells. From this plot we learn that the cluster seems to be elongated in south-east/north-west direction whereas some regions of high galaxy number density show an orientation from the cluster centre along a perpendicular axis, i.e. north-east/south-west. Note that this density map has been computed from the spectroscopic sample alone. Due to the limitations of maximum fibre density, it does not necessarily represent the distribution of the underlying total galaxy population. \\   
   
\subsection{Local fractions of surface number densities}
\begin{figure}
   \centering
   \includegraphics[angle=0,width=\columnwidth]{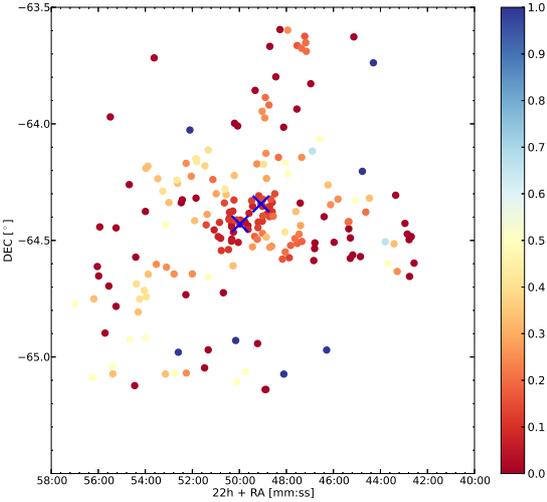}
      \caption{Local fractions of spiral galaxies. Note the deficiency of spirals in the cluster centre and their relatively high abundance at low galaxy number densities. BCG1 and BCG2 are represented by blue crosses. Note that the values associated with the colour bar are fractions of densities and therefore have no unit.
      }
         \label{fig:spir-tot_min}
   \end{figure}

Revisiting the unexpected trends in the radial plots (fractions of EL-galaxies, equivalent widths, star formation rate, $M_{20}$, fractions of no-EL spirals) at distances of 3 Mpc to 4 Mpc from the cluster centre we investigate the local fraction of various morphological galaxy types. Fig. \ref{fig:spir-tot_min} shows the local fraction of spiral galaxies (objects from the spiral sample, see Sec. \ref{susec:morph}) relative to the extended morphological sample (EMS) within a radius of 600 kpc around each cluster member galaxy. This circle size represents the best compromise between spatial resolution and number of galaxies per resolution element. Note that, for the sake of clarity, the symbol sizes do not correspond to a size of 600 kpc but are much smaller. In accordance with previous studies on the abundance of galaxy species with respect to their environment (e.g. \citealt{gallazzi08}) we find a distinct deficiency of spiral galaxies in the cluster centre. Their local fraction reaches values between 0.5 and 0.6 at clustercentric distances greater than $r_{200}$. This behaviour reflects the well-known morphology-density relation found by \citet{dressler80}. 
However, we also find areas with local fractions $>$0.7 to the north-west and to the south-east of our field-of-view, both at a distance between 3 Mpc and 4 Mpc from the cluster centre. We do not find an increased fraction of spiral galaxies in the overdense regions indicated in Fig. \ref{fig:mem_grid_25}.\\    

\begin{figure}
   \centering
   \includegraphics[angle=0,width=\columnwidth]{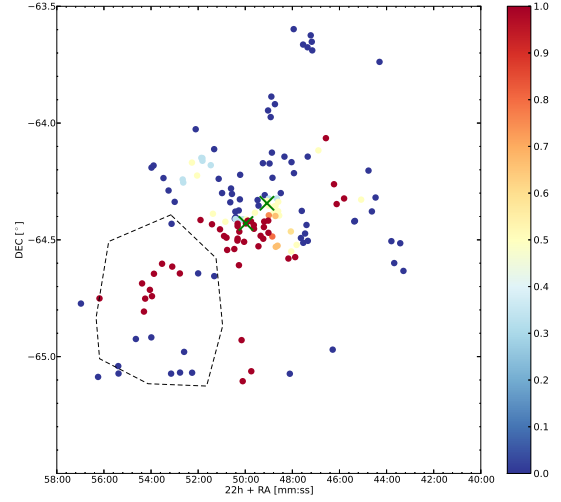}
      \caption{Local fractions of no-EL spiral galaxies (with respect to the spiral sample). Note the relative high abundance of no-EL spirals in areas to the north-west and to the south-east of the cluster centre. BCG1 and BCG2 are represented by green crosses, objects assigned to component C are found within the area enclosed by the dashed lines. Also note that the values associated with the colour bar are fractions and therefore have no unit. The colour bar is inverted with respect to Fig. \ref{fig:spir-tot_min}. 
      }
         \label{fig:noelspir-spir_min}
   \end{figure}

The local fraction of no-EL spirals, i.e. morphologically classified spirals without detectable [OII] or H$\alpha$ emission, with respect to the total spiral sample is depicted in Fig. \ref{fig:noelspir-spir_min}. According to previous results (e.g. \citealt{vogt04,boesch13}) this species of galaxies is expected to be found near the cluster centre for relaxed clusters. However, A3921 shows relatively high local fractions of no-EL spirals not only in its core but also in certain areas found at about 2.9-4.2 Mpc to the north-west and to the south-east of the cluster centre.\\ 
We remind the reader that 'centre' here refers to the position of BCG1 which is also justified by X-ray data on the central 30' of A3921 \citep{belsole05}. These data reveal just a slight shift of 17'' between the peak in X-ray intensity and the position of BCG1. X-ray residuals remaining after the subtraction of a 2D $\beta$-model are found in the vicinity and to the south of BCG2 showing elongations towards BCG1 and peaking to the south-west of BCG2 - thereby giving rise to the hypothesis of an off-axis merger between A3921-A and A3921-B along south-east/north-west direction in the plane of the sky \citep{belsole05}.\\      
   
\begin{figure}
   \centering
   \includegraphics[angle=0,width=\columnwidth]{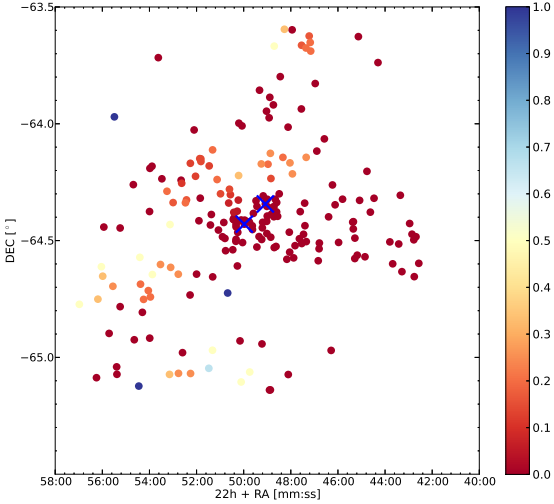}
      \caption{Local fractions of peculiar galaxies. Note the relative high abundance of peculiars in areas to the south-east of the cluster centre as well as the deficiency of peculiars in the western half of our field-of-view. BCG1 and BCG2 are represented by blue crosses. Also note that the values associated with the colour bar are fractions and therefore have no unit.
      }
         \label{fig:pec-tot_min}
   \end{figure}

\begin{figure}
   \centering
   \includegraphics[angle=0,width=\columnwidth]{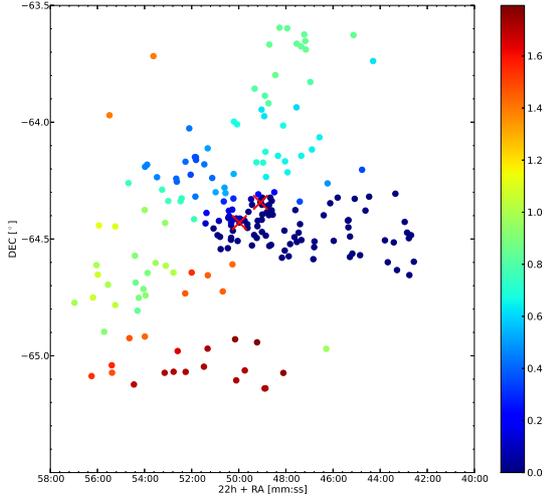}
      \caption{Standard deviation (computed amongst 14 next neighbours, normalised to the standard deviation of the whole EMS) of the local fractions of peculiar galaxies. Note the accumulation of objects with high values in areas to the south-east of the cluster centre. BCG1 and BCG2 are represented by red crosses. The values associated with the colour bar are normalised and therefore have no unit.
      }
         \label{fig:pec-tot_std__min}
   \end{figure}

It can be seen from Fig. \ref{fig:visual} that the morphological classification included peculiar galaxies, i.e. galaxies showing disturbed morphology. Fig. \ref{fig:pec-tot_min} shows the local fraction of peculiar galaxies relative to the extended morphological sample. Peculiars seem to be over-represented in areas to the south-east of the cluster core. In Fig. \ref{fig:pec-tot_std__min} the colour code represents the standard deviation (for a given galaxy and its 14 next neighbours) of the local fraction of peculiars with respect to the extended morphological sample. The values are normalised to the standard deviation of the whole distribution which is $\sim$0.175 (with a mean value of $\sim$0.086).\\
     
\section{Discussion}
Our analyses of spectroscopical and photometric observations of $\sim$200 galaxies in the merging cluster Abell 3921 confirm the presence of two BCGs at a projected separation of $\sim$850 kpc. Each of these giant elliptical galaxies is associated with one of the merging subclusters (A3921-A, A3921-B) as pointed out in \citet{ferrari05}. We also find the subclusters to reside at similar median redshifts of $\sim$0.093 which is in compliance with the scenario of a merger axis almost parallel to the plane of the sky \citep{ferrari05, belsole05}. A Dressler-Shectman test reveals a third cluster component (A3921-C) in the south-east region of A3921-A. Within our sample we find 16 galaxies which are likely to be members of this component. Given its median redshift of 0.086 we argue that A3921-C is falling onto the system A3921-(A+B) from a greater radial distance. Its high fraction of peculiar galaxies yields tentative evidence that A3921-C might be a large galaxy group or even consist of several groups.\\

Our mass estimates for A3921-A and A3921-B exceed the values found by \citet{ferrari05} and \citet{belsole05}. The detection of A3921-C lowers the mass estimate for A3921-A with respect to A3921-B. Hence we find a mass ratio of A3921-B and A3921-A of $\sim$1:2 whereas \citet{ferrari05} find a mass ratio of $\sim$1:5. However, this does not contradict their favoured scenario of an off-axis merger between objects of unequal mass (with A3921-A being the more massive object) in its central phase ($t_{0}\sim\pm$0.3 Gyr where $t_{0}=0$ denotes the time of coalescence). In a follow-up investigation \citet{kapferer06} compare the ICM metallicities of A3921 \citep{belsole05} to numerically simulated cluster mergers. \citet{belsole05} find the highest metallicity in the region of A3921-B and further to the west. Assuming that ejected metals follow behind their ejectors, \citet{kapferer06} hence favour a pre-merger scenario, i.e. $t_{0}\sim$-0.3 Gyr. Our new mass ratio and total mass estimates are in compliance with this scenario.\\

Our estimate for $r_{200}$ of A3921-(A+B) exceeds the values given by \citet{ferrari05} but is in excellent agreement with the results of the caustics method applied to a larger galaxy catalogue of A3921 constructed from literature data. However, our mass estimate clearly exceeds the caustic results. This discrepancy certainly is due to the fact that our mass estimates are based on the assumption of a virialised galaxy population which might not be fulfilled for A3921.\\

The radial trends in fractions of EL-galaxies, equivalent widths, star formation rates, concentration index, Gini coefficient and $M_{20}$ up to clustercentric distances of $\sim$$r$$_{200}$ correspond to the behaviour suggested by the morphology-density relation. Fractions of EL-galaxies (and EL-spirals), equivalent widths, star formation rates and $M_{20}$ surprisingly show a decrease (in absolute numbers) at $\sim$3.5 Mpc clustercentric distance. The reason for these decreases are high fractions of no-EL spirals at radial distances of 3 Mpc to 4 Mpc in regions to the north-west and south-east of the cluster centre. Note that this result is robust against the definition of cluster membership: even if we restrict our analysis to virialised cluster members according to the caustics approach, the subsample of no-EL spirals is reduced by only one object (whereas the total sample of cluster members is reduced by 96 galaxies).\\

The highest fractions of no-EL spirals are, however, found in the cluster core at radial distances smaller than 1.2 Mpc. Analysing a volume-limited sample of SDSS data, also e.g. \citet{goto03} find that no-EL spirals (or "passive" spirals) mainly occur in high-density environments. Regarding the high probability of dynamical interaction processes (e.g. galaxy mergers or galaxy-galaxy tidal interactions) to disturb or destroy spiral arms \citet{goto03} propose a cluster related mechanism to explain their findings. According to more recent results (e.g. \citealt{boesch13}) no-EL spirals are expected to occur near the cluster centre for relaxed clusters. The generally adopted scenario is that no-EL spirals close to the cluster centre have been subject to ram pressure stripping during their infall. \citet{quilis00} show that on a time scale of $\sim$100 Myr ram pressure stripping may remove the majority of a galaxy's atomic hydrogen content, thereby quenching star formation. \citet{vogt04} argued that no-EL spirals (or "quenched" spirals) could represent an intermediate stage of a morphological transformation of spiral galaxies (falling into the cluster from the field) into S0s. In their extensive analysis of two  virialised clusters at z$\sim$0.5 \citet{moran07} address the question of a transition from spiral galaxies into S0s. They find the highest fractions of passive spirals and S0s close to the cluster centres but also observe both species to reside in infalling groups out to clustercentric radii beyond $r_{200}$. The authors conclude that a combination of ram pressure stripping and galaxy-galaxy tidal interactions is required to explain their observations. The resolution of our imaging data is, however, insufficient to identify S0 galaxies. Nevertheless, the large fractions of no-EL spirals observed at small clustercentric radii in Abell 3921 are in compliance with the transformation scenario.\\ 
 
No-EL spirals that occur at clustercentric distances greater than 3 Mpc are found in the directions of A3921-B to the north-west and A3921-C to the south-east. This explains why we find decreases at 3-4 Mpc in the aforementioned radial trends not only in the total cluster sample but also in subsamples A3921-A and A3921-B. The occurrence of no-EL spirals at large clustercentric radii could be explained by shocks in the ICM due to the ongoing cluster merger. A merger shock front represents a steep increase in local ICM density, pressure and temperature. A shock front moving through a galaxy (or vice versa) hence could be the site for an increased ram pressure effect. Ram pressure stripping can increase the SFR on short timescales (e.g. \citealt{steinhauser12}) and decrease it on longer timescales due to gas consumption (e.g. \citealt{quilis00}). Likewise, the influence of a shock front probably leads to a SFR decrease on long timescales. \citet{chung09} investigated the 6.2$\mu$m polycyclic aromatic hydrocarbon emission feature of galaxies behind and in front of a merger shock front in the "Bullet cluster". In particular, they focussed on the flux ratio of the 8$\mu$m band (containing the 6.2$\mu$m feature at the observed redshift) and the 4.5$\mu$m band which is a proxy for stellar mass. The $m_{4.5} - m_{8}$ colour was hence used as a proxy of specific SFR. \citet{chung09} find that post-shock galaxies have on average a lower specific SFR than their pre-shock counterparts. Applying a standard KS test they find that the difference in colour between both samples is approximately of $2\sigma$. They conclude that a shock front induced by a merger event as represented by the bullet cluster does not drastically quench star formation. However, the observed difference between galaxies in pre- and post-shock regions is significant. Evaluating data on the major merger Abell 2744, \citet{owers12} find very bright star forming knots and filaments in the stripped wake of four infalling spiral galaxies. Three of these so-called "jellyfish" objects are found close to merger-related features in the ICM. Hence \citet{owers12} suggest that the star formation in the stripped wakes has been triggered by the increase of pressure during an interaction with a merger shock. They consider their findings as evidence for the accelerated morphological transformation of galaxies due to shock interaction. As observations of A3921 in radio (ATCA, \citealt{ferrari06}) and X-ray (XMM-Newton, \citealt{belsole05}) did only cover regions at clustercentric distances $\lesssim$ 15' without detecting any traces of shock fronts, our hypothesis of shocks as the reason for quenched spirals at large clustercentric radii is, to some extent, speculative. However, the existing radio and X-ray data did \textit{not} cover radii out to $\sim$3-4 Mpc where we observe the quenched spirals.

\section{Conclusions}
   
In summary, our spectroscopical and morphological analyses motivate the following conclusions:

\begin{itemize}
\item
Component C might be a massive group falling onto the cluster from the south-east (projected onto the plane of the sky) with a high peculiar line-of-sight velocity towards the observer. The high fraction (25$\%$) of peculiar galaxies in component C would be very unusual for a cluster.
\item
Component C might be further extended in westward direction than indicated by the DS-test. This is suggested by e.g. the widespread distribution of peculiar galaxies in the respective region.
\item
The clustercentric distances and the locations in z-space of the areas north-west and south-east of the cluster centre showing overdensities of spirals, no-EL spirals and peculiar galaxies suggest that the merger between A3921-A and A3921-B is dynamically more advanced than the merger of A3921-(A+B) and component C.
\item
The high fraction of no-EL spirals in the cluster core supports the hypotheses of morphological transformation in regions of increased density. According to this scenario no-EL spirals represent a transition stage between field spirals and cluster S0s.
\item
The unexpected occurrence of no-EL spirals at distances $>$2$r_{200}$ could at least partially be explained by shock fronts in the ICM due to the ongoing cluster merger.
\end{itemize}

\begin{acknowledgements}
We thank the referee, Florence Durret, for her constructive comments which helped a lot to improve the manuscript. We warmly thank all the AAO team and in particular Rob Sharp and Fred Watson for having performed and pre-reduced our 2dF observations. We are very grateful to Tom Mauch for having provided the code \texttt{runz}. Florian Pranger, Asmus B\"{o}hm and Sabine Schindler are grateful for funding by the Austrian Funding Organisation FWF through grant P23946-N16. Chiara Ferrari acknowledges financial support by the "\textit{Agence Nationale de la Recherche}" through grant ANR-09-JCJC-0001-01. Antonaldo Diaferio acknowledges partial support from the INFN grant PD51.
\end{acknowledgements}    

\bibliographystyle{aa}
\bibliography{flo_bib}

\end{document}